\documentclass[useAMS,usenatbib]{mn2e}

\usepackage{graphicx}

\newcommand{\Msun}{M_\odot}

\newcommand{\Mstar}{M_\star}

\newcommand{\Zsun}{Z_\odot}
\newcommand{\Zcrit}{Z_{\rm crit}}
\newcommand{\Zgal}{Z_{\rm gal}}

\title[Revisiting Metallicity of GRB Hosts]
{Revisiting Metallicity of Long Duration Gamma-Ray Burst Host Galaxies: 
The Role of Chemical Inhomogeneity within Galaxies}

\author[Y. Niino]{Y. Niino$^{1,2}$\thanks{E-mail: yuu.niino@nao.ac.jp}\\
 $^{1}$Division of Optical \& Infrared Astronomy, National Astronomical Observatory of Japan, 
2-21-1 Osawa, Mitaka, Tokyo 181-8588, Japan\\
 $^{2}$Department of Astronomy, School of Science, Kyoto University, 
Kitashirakawa-Oiwakecho, Sakyo-ku, Kyoto 606-8502, Japan}

\begin{document}

\date{}

\pagerange{\pageref{firstpage}--\pageref{lastpage}} \pubyear{2011}

\maketitle

\label{firstpage}

\begin{abstract}
We predict the metallicity probability distribution function (PDF) 
of long gamma-ray burst (GRB) host galaxies at low-redshifts ($z \la 0.3$) 
when GRBs occur only in low-metallicity environment, 
assuming empirical formulations of galaxy properties. 
We discuss contribution of high-metallicity galaxies 
to the cosmic rate of low-metallicity GRBs, taking
internal dispersion of metallicity within each galaxy into account. 
Assuming GRBs trace low-metallicity star formation $< \Zcrit$: 12+log$_{10}$(O/H) = 8.2, 
we find that GRB host galaxies may have systematically higher-metallicity than that of GRB progenitors. 
Furthermore, we expect $\ga$ 10\% of the host galaxies to have 12+log$_{10}$(O/H) $> 8.8$, 
if galaxies have internal dispersion of metallicity comparable to that observed in the Milky Way. 
Our results show that the low-metallicity scenario of GRB progenitors 
can be reconciled with the recent discoveries of the high-metallicity host galaxies of GRBs. 
We also show possible bimodality in the host metallicity PDF 
that results from the single progenitor model of GRBs. 
If found in future observation, the bimodality can be a clue to 
constrain the nature of GRB progenitors. 
\end{abstract}


\section{Introduction}

It is now broadly agreed that some of long gamma-ray bursts (GRBs) 
originate in core-collapse supernovae (CC SNe). 
However, not all CC SNe produce GRBs, 
and the criteria that discriminates GRBs from general CC SN is 
one of the most outstanding questions about GRBs. 

Some theoretical studies on the origin of GRBs 
using stellar evolution models suggest that a low-metallicity 
may be a necessary condition for a GRB to occur 
\citep[$Z < $ a few $\times\ 0.1\Zsun$, e.g.,][]{MacFadyen:99a,Yoon:05a,Yoon:06a,Woosley:06a}. 
It has also been suggested from observations 
that metallicity distribution 
of GRB host galaxies at redshift $z < 0.25$ is significantly biased 
towards low metallicities compared to the expectation 
when GRBs are unbiased tracers of star formation \citep{Stanek:06a,Modjaz:08a}. 

Furthermore, some observations suggest that the GRB host galaxies are 
systematically fainter and smaller than those of the core-collapse SNe 
\citep{Le-Floch:03a, Fruchter:06a, Svensson:10a}, 
indicating that the GRBs may preferentially occur 
in low metallicity environment, because fainter and lower mass 
galaxies generally have lower metallicities.  
These interpretations have also been supported by other theoretical studies 
using the models of galaxies \citep[e.g.][]{Wolf:07a, Niino:11a}. 

However, recent discoveries of high-metallicity host galaxies of some GRBs 
cast doubt on the requirement of low-metallicity in GRB occurrence 
\citep{Levesque:10d, Levesque:10f, Han:10a, Hashimoto:10a}. 
On the other hand, it should be kept in mind that GRB host galaxies 
may have different metallicity from that of GRB progenitors, 
because a galaxy is not a chemically homogeneous object. 
To decide whether the discoveries of the high-metallicity host galaxies 
are consistent to the low-metallicity requirement or not, 
we need to quantitatively consider contribution of high-metallicity galaxies 
to the cosmic rate of low-metallicity star formation. 

In this study, we predict the metallicity probability distribution function (PDF) 
of GRB host galaxies at $z \la 0.3$, 
assuming the low-metallicity requirement and 
taking the metallicity dispersion within each galaxy into account. 
We discuss whether the expected rate of low-metallicity GRBs 
in high-metallicity galaxies is significant to explain the observations or not. 
We use empirical formulations of galaxy properties, 
and assume GRBs trace low-metallicity star formation $< \Zcrit$: 12+log$_{10}$(O/H)$= 8.2$. 
In \S~\ref{sec:model}, we describe empirical formulations of galaxy properties 
and GRB rate models which we use in this study. 
In \S~\ref{sec:result}, we show the expected metallicity and mass distributions 
of GRB host galaxies and discuss their implications. 
In \S~\ref{sec:Z-SFR}, we also discuss how the expected metallicity distribution changes 
if there is a correlation between metallicity and star formation rate of a galaxy, 
as it is claimed in \citet{Mannucci:10a}. 
We summarize our conclusion in \S~\ref{sec:discussion}. 

\section{Models}
\label{sec:model}
\subsection{Galaxy Properties}
\label{sec:galaxy}
To compute the expected metallicity PDF of GRB host galaxies at $z \la 0.3$, 
we assume empirical formulation of stellar mass function, 
mass--star formation rate ($\Mstar$--SFR) relation of galaxies, 
and mass--metallicity ($\Mstar$--$\Zgal$) relation of low-redshift galaxies. 
Our approach is similar to that of \citet{Stanek:06a}, 
who have calculated expected metallicity PDF of GRB host galaxies 
when GRBs trace star formation without any metallicity dependence. 
However, we step further to include the low-metallicity preference of GRBs 
considering the chemical inhomogeneity within galaxies. 

It should be noted that $\Mstar$ is calibrated using 
different initial mass functions (IMFs) in different studies. 
In this study, we assume conversion among stellar mass scales for different IMFs 
as: $M_{\rm Salpeter} = 1.43\times M_{\rm diet Salpeter}$
 = $1.5\times M_{\rm Kroupa} = 1.8\times M_{\rm Chabrier} = 1.8\times M_{\rm BG03}$ 
\citep{Salpeter:55a, Bell:01a, Kroupa:01a, Chabrier:03a, Baldry:03a}. 
Hereafter stellar masses are scaled for Salpeter IMF, unless otherwise stated. 

\begin{figure}
 \includegraphics[width=84mm]{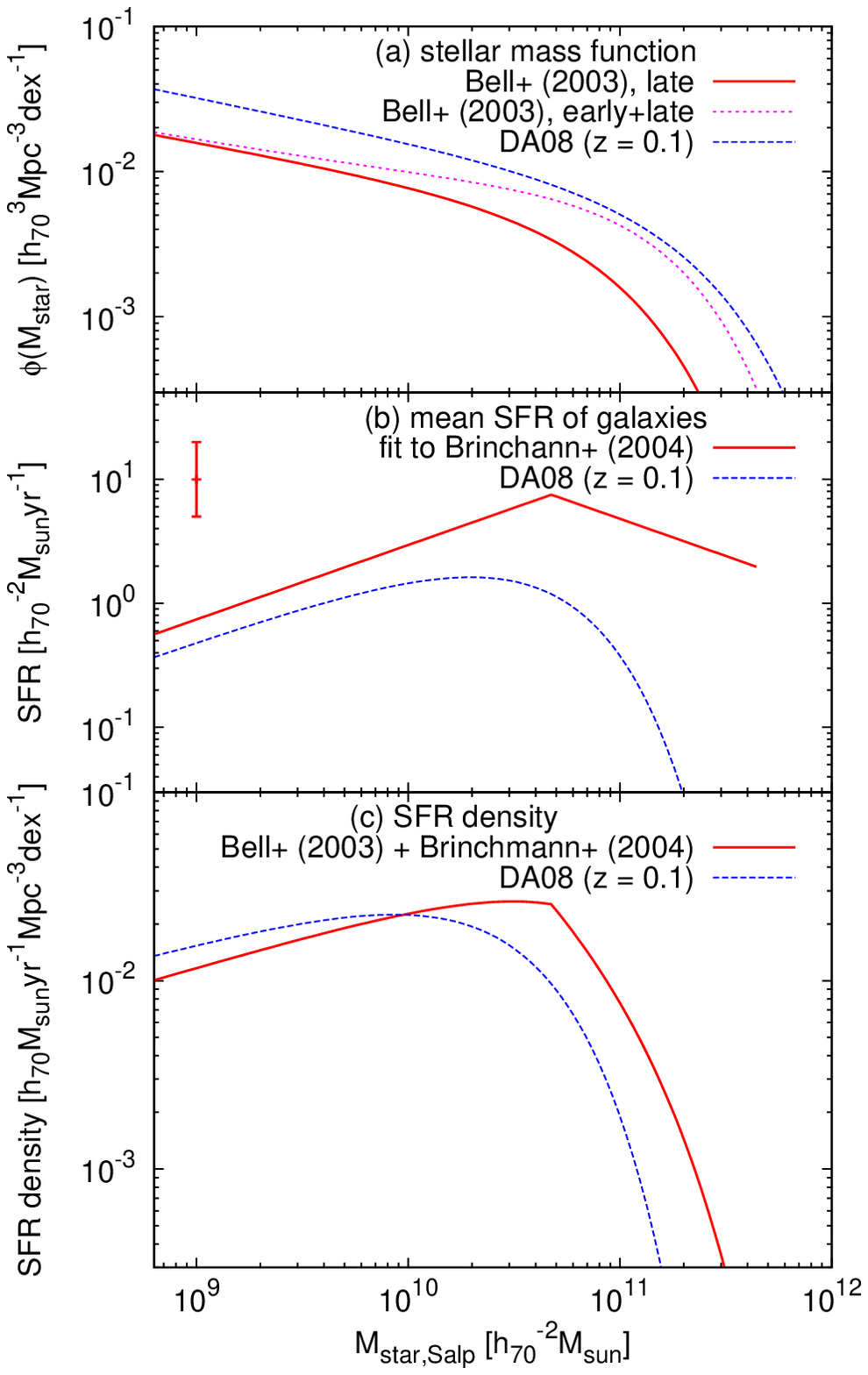}
 \caption{
 Empirical formulations of stellar mass function, $\Mstar$--SFR relation, 
 and SFR density for late-type galaxies \citep{Bell:03a, Brinchmann:04a}. 
 Those for all types of galaxies (DA08) are plotted together. 
 In panel (b), the typical dispersion of galaxy SFR around the relation is shown.
 }
 \label{fig:MSFR}
\end{figure}

We use empirical formulations of the stellar mass function \citep{Bell:03a}, 
and the $\Mstar$--SFR relation [\citet{Stanek:06a}, a fit to the observation by \citet{Brinchmann:04a}] 
of low-redshift ($z \la 0.3$) late-type galaxies. 
We assume the dispersion of the $\Mstar$--SFR relation 
to be $\sim 0.3$ dex following \citet{Stanek:06a}. 
Using the mass function and the $\Mstar$--SFR relation, 
we compute cosmic SFR density as a function 
of $\Mstar$: $\rho_{\rm SFR}(\Mstar)$ [$\Msun$yr$^{-1}$Mpc$^{-3}$dex$^{-1}$]. 
These models are shown in Fig.~\ref{fig:MSFR}. 
We only consider galaxies with log$_{10}$~$\Mstar > 8.0$, 
which corresponds to the lowest-mass of GRB host galaxies ever known. 
Both of the stellar mass function and the $\Mstar$--SFR relation 
may suffer from selection effects of the galaxy sample. 
To demonstrate the uncertainty, we also use $\rho_{\rm SFR}(\Mstar)$ 
derived from observation of galaxies of all-type 
\citep[][DA08]{Drory:08a}. 

Various methods have been proposed to measure gas metallicity of galaxies, 
but they do not always agree with each other \citep[e.g.,][]{Kennicutt:03a, Kewley:08a}. 
When we discuss metallicity of galaxies, we use metallicities calibrated  
with \citet[hereafter KK04]{Kobulnicky:04a} method in this study. 
We use the $\Mstar$--$\Zgal$ relation with the KK04 calibration 
presented in Eq. 8 of \citet[hereafter S05 relation]{Savaglio:05a}. 
We assume the dispersion of the relation 
to be 0.1 dex \citep{Tremonti:04a}. 

When we consider the dispersions of $\Mstar$--SFR relation and $\Mstar$--$\Zgal$ relation, 
we assume that the offset from $\Mstar$--$\Zgal$ relation 
and that from $\Mstar$--SFR relation are independent of each other. 
This assumption is supported by the observed no correlation 
between H$\alpha$ equivalent width and SFR \citep{Tremonti:04a}. 
However, it is also claimed that those offsets are correlated \citep{Mannucci:10a}, 
and we discuss the case where of the offsets are correlated in \S~\ref{sec:Z-SFR}. 

\subsection{GRB Rate and Internal Dispersion of Metallicity within Each Galaxy}
\label{sec:internal}
Observations of nearby galaxies, 
including the Milky Way (MW) and the Magellanic clouds, 
show that the galaxies have internal dispersion of metallicity within them 
\citep[$\sim 1$ dex in MW and $\sim 0.3$ dex in the Large Magellanic Cloud (LMC), e.g.,]
[]{Rolleston:00a, Rolleston:02a}. 
Furthermore, there is a $\sim 0.4$ dex variation of 12+log$_{10}$(O/H) 
among HII regions in the host galaxy of GRB 980425/SN 1998bw 
which is comparable to $3\sigma$ error of the metallicity calibration \citep{Christensen:08a}. 
To demonstrate effects of the chemical inhomogeneity, 
we assume metallicity of SFR in a galaxy 
has a log-normal distribution with dispersion $\sigma_{Z, {\rm int}}$ around $\Zgal$, 
although metallicity distribution of star forming gas within a galaxy is hardly understood. 

\begin{figure} 
\includegraphics[width=84mm]{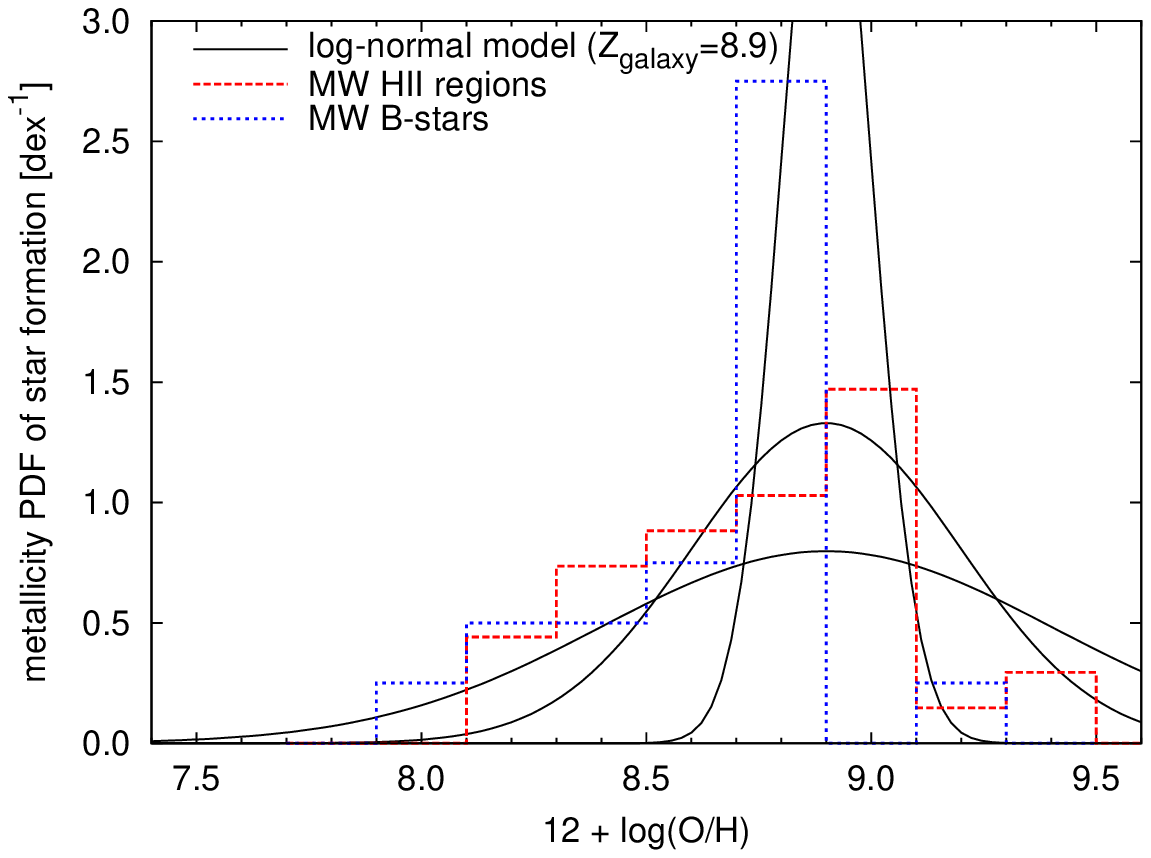}
 \caption{
  The log-normal models of the internal metallicity dispersion 
  within a galaxy is shown ($\sigma_{Z, {\rm int}} = 0.1, 0.3$ and 0.5, solid curves), 
  together with the metallicity distributions of the HII regions and the young B-type stars 
  \citep[][red dashed and blue dotted, respectively]{Afflerbach:97a, Rolleston:00a}. 
  We note that the metallicities measured in \citet{Afflerbach:97a} 
  and \citet{Rolleston:00a} may be inconsistent with KK04 metallicities. 
 } 
 \label{fig:Zdisp}
\end{figure}

In Fig.~\ref{fig:Zdisp}, we plot the log-normal models with $\Zgal=8.9$ 
and $\sigma_{Z, {\rm int}} = 0.1, 0.3$ and 0.5. 
For comparison, we plot the observed metallicity distribution 
of HII regions and young B-type stars in the MW 
\citep{Afflerbach:97a, Rolleston:00a} together, 
although they do not necessarily represent 
overall metallicity PDF of star forming gas in the MW. 
Both of the HII regions and the B-type stars 
typically have 12+log$_{10}$(O/H) $\sim 8.9$, 
while $\sim$ 10\% have 12+log$_{10}$(O/H) $\la 8.2$,
which is roughly comparable to the case of $\sigma_{Z, {\rm int}} = 0.5$. 
Although the models and the observation 
do not agree with each other in the high-metallicity end, 
we note that the high-metallicity end 
of the internal metallicity distribution has small effect on our results. 

In this study, we assume GRB rate is proportional 
to low-metallicity SFR with $Z < \Zcrit$, 
and express $R_{\rm GRB} \propto \epsilon_{\rm GRB}(\Zgal)\times$SFR, 
where $\epsilon_{\rm GRB}(\Zgal) \equiv {\rm SFR}_{Z<\Zcrit}/{\rm SFR}$ is a GRB efficiency in a galaxy. 
We do not need to determine the proportionality constant, 
because our interest is in the metallicity PDF of GRB host galaxies. 
Under the assumption of the log-normal metallicity distribution, 
the GRB efficiency can be written as: 
\begin{equation} 
	\epsilon_{\rm GRB}(\Zgal) 
	 = 0.5 \times {\rm erfc}[\frac{-{\rm log}_{10}(\Zcrit/\Zgal)}{\sqrt{2\sigma_{Z, {\rm int}}}}]. 
\label{eq:GRBeff} 
\end{equation} 
Here we assume $\Zcrit$ to be 12+log$_{10}$(O/H) = 8.2, 
which roughly corresponds to 0.2--0.4$\Zsun$. 
The expected metallicity distribution of GRB host galaxies 
is $\hat{\rho}_{\rm GRB}(\Zgal) \propto \epsilon_{\rm GRB}(\Zgal)\hat{\rho}_{\rm SFR}(\Zgal)$, 
where $\hat{\rho}_{\rm SFR}(\Zgal)$ is the $\rho_{\rm SFR}(\Mstar)$ 
projected to $\Zgal$ axis using the $\Mstar$--$\Zgal$ relation. 

It should be noted that \citet{Kocevski:09a} performed similar investigation 
to that in this study to calculate the mass PDF 
of GRB host galaxies when GRBs trace low-metallicity SFR, 
and claimed $\Zcrit > 0.5\Zsun$ which is contrary to our assumption. 
However, they assumed that $R_{\rm GRB} = 0$ when $\Zgal > \Zcrit$ 
without taking the internal dispersion of galaxies into account, 
and their conclusion might be affected by this assumption. 

\section{The Metallicity PDF of GRB Host Galaxies}
\label{sec:result}
\begin{figure*}
 \includegraphics[scale=0.6]{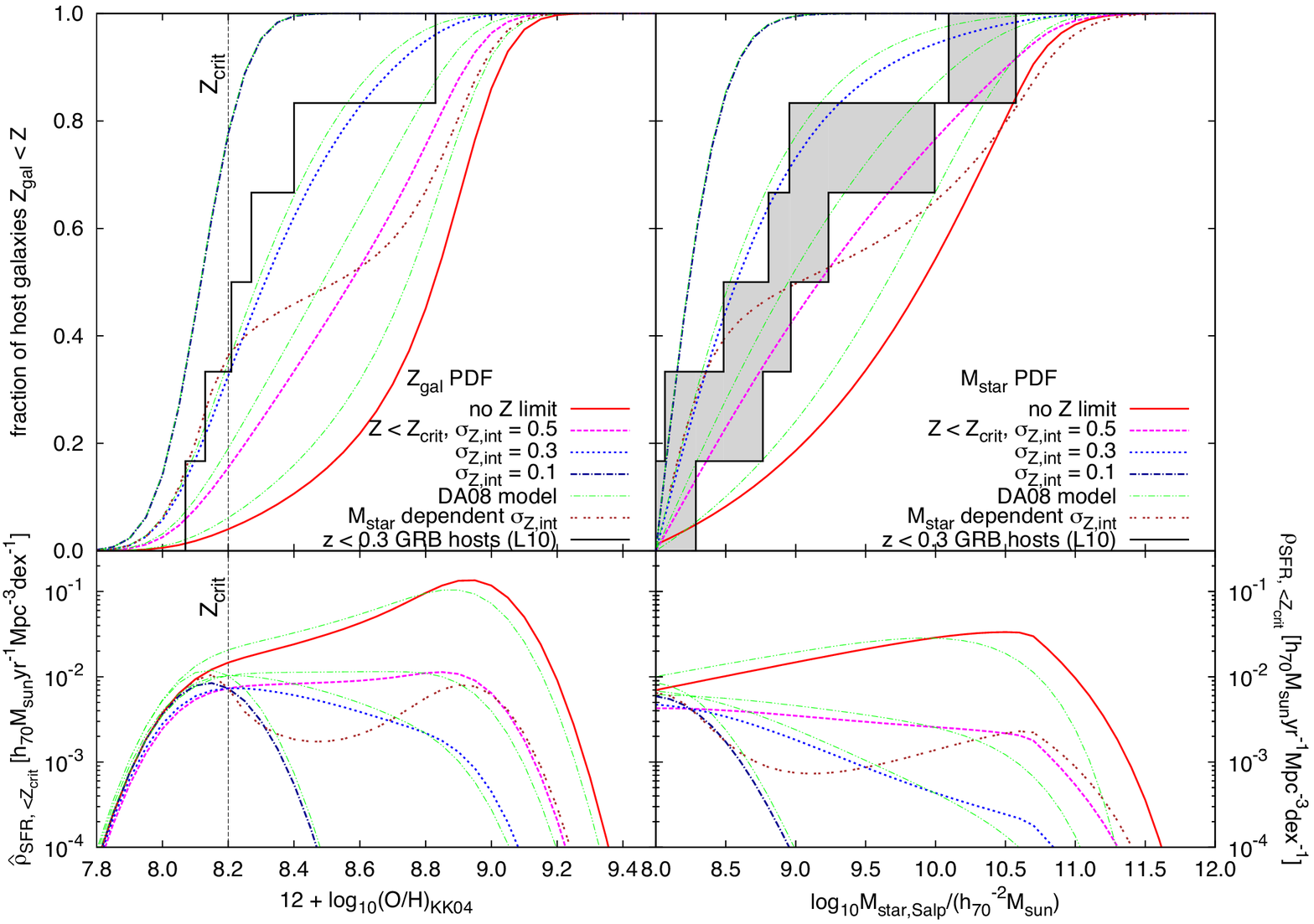}
 \caption{
 The $\Zgal$ PDF ({\it left panels}) and $\Mstar$ PDF ({\it right panels}) 
 of GRB host galaxies predicted in our model. 
 The {\it top panels} show the normalized PDF, 
 while the {\it bottom panels} show the $R_{\rm GRB}$ 
 weighted metallicity function of galaxies. 
 Observed GRB host galaxies at $z < 0.3$ \citep[][L10]{Levesque:10f} are shown together. 
 The PDFs predicted without the effect of $\Zcrit$ are plotted with red solid lines, 
 while results with $\sigma_{Z, {\rm int}}$ = 0.5, 0.3 and 0.1 dex are shown 
 with dashed (magenta), dotted (blue), and dot-dashed (dark blue) lines, respectively. 
 The green double-dotted dashed line represents the results with DA08 models. 
 The results in the case where $\sigma_{Z, {\rm int}}$ 
 correlates with $\Mstar$ are shown with double-dotted line (brown). 
 }
 \label{fig:main}
\end{figure*}

The predicted metallicity PDF of GRB host galaxies 
is shown in the left panels of Fig.~\ref{fig:main}. 
The model without metal cutoff (i.e. $\epsilon_{\rm GRB}(\Zgal) = 1.0$) 
is consistent to the results of \citet{Stanek:06a}, 
and it shows that more than 50\% of low-redshift star formation takes place in 
high-metallicity galaxies with 12+log$_{10}$(O/H) $> 8.8$, 
which is much higher fraction than high-metallicity galaxies in observed GRB host galaxies. 

Now we consider the effect of the metal cutoff on the metallicity distribution of the host galaxies. 
The contribution of $\Zgal > \Zcrit$ galaxies is not zero due to the effect of the internal dispersion. 
The results with $\sigma_{Z, {\rm int}} = 0.1, 0.3$ and 0.5 are shown in Fig.~\ref{fig:main}. 
In the cases of $\sigma_{Z, {\rm int}} \geq$ 0.3 dex, 
more than 50\% of GRB host galaxies have $\Zgal > \Zcrit$, 
suggesting that the progenitor metallicity can be systematically different from the host metallicity. 

The contribution of the high-metallicity galaxies (12+log$_{10}$(O/H) $> 8.8$) 
to the cosmic GRB rate is equivalent to that of $\Zgal < \Zcrit$ galaxies
when they have $\sigma_{Z, {\rm int}} =$ 0.5 dex. 
In the case of $\sigma_{Z, {\rm int}}$ = 0.5 (0.3) dex, 
roughly 25\% (5\%) of the host galaxies have 12+log$_{10}$(O/H) $> 8.8$, 
suggesting the hypothesis that GRBs occur only in low-metallicity environment 
does not contradict to the recent observations of high-metallicity host galaxies of GRBs. 
We note that the prediction of the DA08 model 
is not largely different from that of the late-type galaxy model. 
The expected mass PDF of GRB host galaxies is shown 
in the right panels of Fig.~\ref{fig:main} in the similar manner to the metallicity PDFs. 

We have assumed that $\sigma_{Z, {\rm int}}$ 
is same among all galaxies in 
discussions above. 
However, the internal metallicity dispersion within galaxies is not well known, and 
it is possible that galaxies with different $\Mstar$ (or $\Zgal$) typically have different $\sigma_{Z, {\rm int}}$. 
In fact, the LMC has smaller internal dispersion of metallicity 
than that in MW \citep[e.g.,][]{Rolleston:02a, Cioni:09a}. 
In that case, the expected metallicity distribution would be different from those discussed above. 
However, the relative contribution of 12+log$_{10}$(O/H) $> 8.8$ galaxies 
compared to the $Z < \Zcrit$ galaxies would be similar to the case of $\sigma_{Z, {\rm int}} \geq$ 0.3 dex 
when high-metallicity galaxies have $\sigma_{Z, {\rm int}}$ compareble to that observed in the MW, 
regardless of $\sigma_{Z, {\rm int}}$ in 12+log$_{10}$(O/H) $\leq 8.8$ galaxies. 

To demonstrate a case in which $\sigma_{Z, {\rm int}}$ correlates with $\Mstar$, 
we employ a toy model of $\Mstar$ dependent $\sigma_{Z, {\rm int}}$, 
\begin{equation}
\sigma_{Z, {\rm int}}(\Mstar) = 0.2\times({\rm log}\Mstar - 8.0). 
\label{eq:sigma-M}
\end{equation}
The results are shown in Fig.~\ref{fig:main}. 
It is interesting that the $\Zgal$ PDF has multi-peak distribution in this model, 
although we consider only one population of GRB progenitors. 

This bimodality can be explained as follows. 
In the case of $\sigma_{Z, {\rm int}} =$ 0.5 dex, 
 $\hat{\rho}_{\rm GRB}(\Zgal)$ is approximately constant 
between 12+log$_{10}$(O/H) $= 8.2$ and 9.0 (see left bottom panel of Fig.~\ref{fig:main}). 
If (1) high-metallicity galaxies (12+log$_{10}$(O/H) $\ga 9.0$) 
typically have $\sigma_{Z, {\rm int}} \sim$ 0.5 dex 
and (2) $\sigma_{Z, {\rm int}}$ is positively correlated with $\Zgal$, 
galaxies with 12+log$_{10}$(O/H) $\la 8.2$ or $\ga 9.0$ 
have similar $\hat{\rho}_{\rm GRB}(\Zgal)$ to that in the case of $\sigma_{Z, {\rm int}} =$ 0.5 dex, 
while smaller $\epsilon_{\rm GRB}(\Zgal)$ in $8.2 <$ 12+log$_{10}$(O/H) $< 9.0$ 
produce a dip in the metallicity PDF between 12+log$_{10}$(O/H) $= 8.2$ and $9.0$. 
Note that $\epsilon_{\rm GRB}$ and $\sigma_{Z, {\rm int}}$ 
are positively correlated when $\Zgal > \Zcrit$. 
Thus it is possible that bimodal distribution of $\Zgal$ 
appears from a single population of GRB progenitors, 
when the conditions (1) and (2) are fulfilled. 
We should keep in mind that bimodality of GRB host galaxy population 
does not necessarily mean bimodal nature of GRB progenitors. 

\section{Correlation beween $\Zgal$ and SFR}
\label{sec:Z-SFR}
In the previous sections, the dispersions of the $\Mstar$--$\Zgal$ relation 
and the $\Mstar$--SFR relation are treated independently. 
However, it is recently claimed that galaxies with higher-SFR 
tend to have lower-$\Zgal$ compared to lower-SFR galaxies 
with similar $\Mstar$ \citep{Mannucci:10a}. 
\citet{Mannucci:11a} and \citet{Kocevski:10a} have investigated 
the effect of the SFR--$\Zgal$ correlation 
on the $\Mstar$--$\Zgal$ relation of GRB host galaxies. 

\begin{figure} 
\includegraphics[width=84mm]{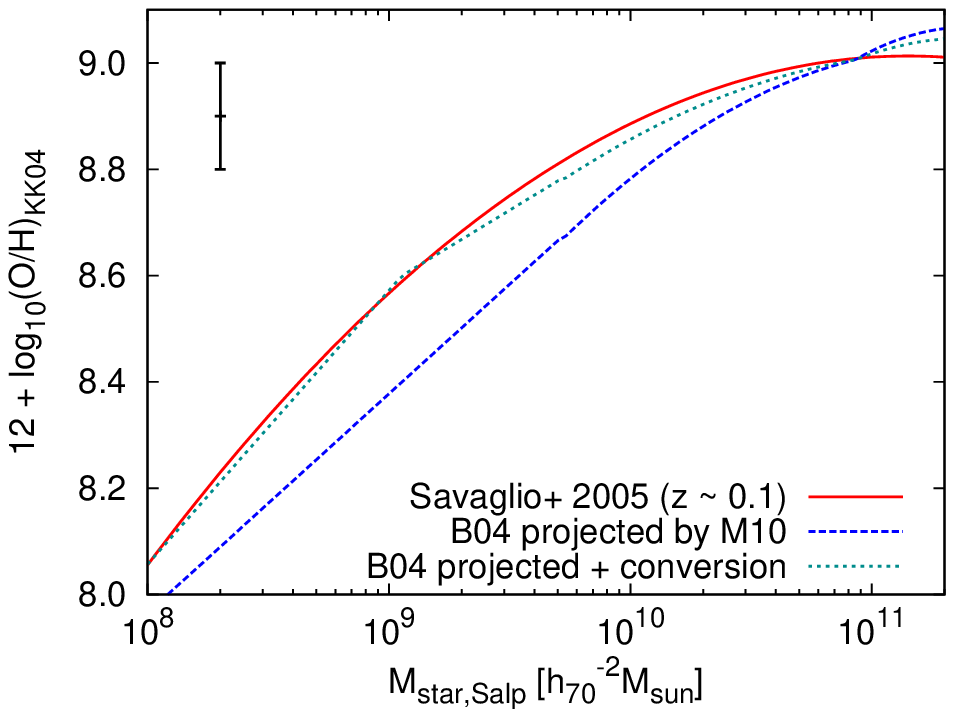}
 \caption{
  The $\Mstar$--$\Zgal$ relation of low-redshift galaxies ($z\la0.3$) 
  presented in \citet[S05 relation, the red solid line]{Savaglio:05a} is compared 
  with the mass-SFR relation \citep{Brinchmann:04a} projected 
  by M11 relation \citep[][blue dashed line]{Mannucci:11a}.
  The conversion of the projected $\Mstar$--SFR relation 
  with Eq.~\ref{eq:N06toKK04} is shown with a cyan dotted line.
 } 
 \label{fig:MZrel}
\end{figure}

In this section, we use $\Mstar$--SFR--$\Zgal$ relation 
\citep[][hereafter M11 relation]{Mannucci:11a} instead of S05 relation, 
to investigate how the metallicity PDF of the host galaxies is changed if there is a SFR--$\Zgal$ correlation. 
It should be noted that M11 relation 
studied in \citet{Mannucci:10a, Mannucci:11a} 
is calibrated with \citet[][hereafter N06]{Nagao:06a} method. 
Hence we can not directly compare predictions of M11 relation 
with predictions and/or observations with KK04 calibration. 
We project the mass-SFR relation described in \S~\ref{sec:galaxy} 
to $\Mstar$--$\Zgal$ plane using M11 relation, 
and compare it with S05 relation. 
In Fig.~\ref{fig:MZrel}, one sees discrepancy between the two relation. 

To make consistent comparison between the two models, 
we assume ad hoc conversion between the two metallicity calibration: 
\begin{eqnarray} 
\alpha_{\rm KK04} = \left \{
\begin{array}{l} 
  \alpha_{\rm N06} + 0.25(\alpha_{\rm N06}-8.0) + 0.1 \ \ (\alpha_{\rm N06} < 8.4) \\
   \alpha_{\rm N06} - 0.33(\alpha_{\rm N06}-8.4) + 0.2 \ \ (\alpha_{\rm N06} \geq 8.4)
\end{array} \right. ,
\label{eq:N06toKK04} 
\end{eqnarray} 
where $\alpha$ represents 12+log$_{10}$(O/H). 
Converted with Eq.~\ref{eq:N06toKK04}, 
the projected $\Mstar$--SFR relation agrees 
with S05 relation in 0.04 dex (Fig.~\ref{fig:MZrel}). 

\begin{figure*} 
\includegraphics[scale=0.6]{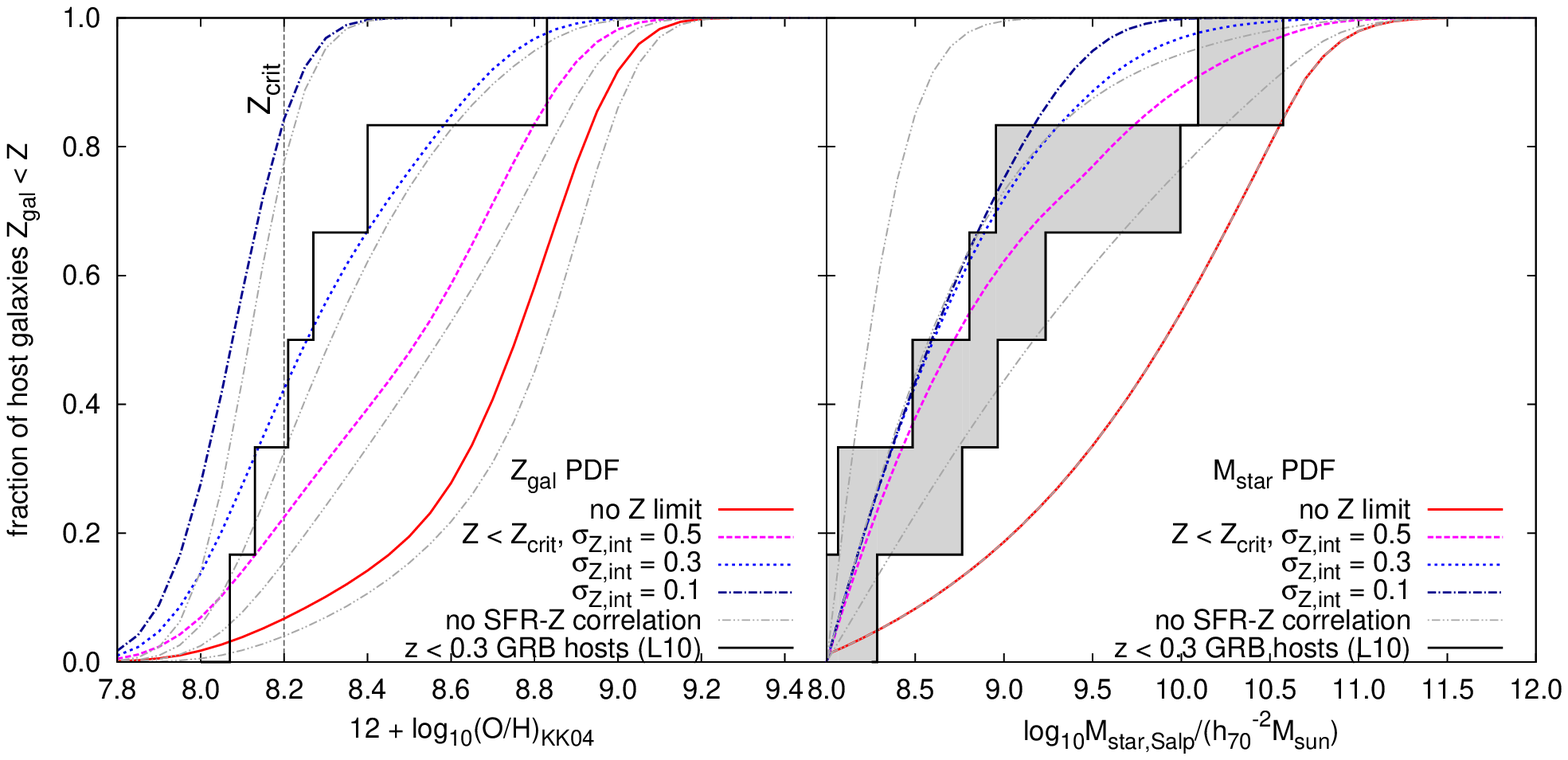}
 \caption{
  Same as the top panels of Fig.~\ref{fig:main}, 
  but M11 relation is assumed instead of S05 relation. 
  The results with S05 relation, 
  which have been shown in Fig.~\ref{fig:main}, 
  are plotted together (grey double-dotted dashed lines). 
 } 
 \label{fig:MSFRZ}
\end{figure*}

The metallicity and mass PDFs predicted 
using M11 relation are shown in Fig.~\ref{fig:MSFRZ}. 
Although the host galaxies have lower-metallicity by $\la$ 0.1 dex 
compared to the case of S05 relation depending on $\sigma_{Z, {\rm int}}$, 
M11 relation alone does not make the matellicity PDF 
consistent to the current sample of GRB hosts without further metallicity effect. 
The metallicity PDFs with the metallicity cut-off may agree 
with the observations, as well as in the case of S05 relation. 

The predicted host galaxies in the case of $\sigma_{Z, {\rm int}} = 0.5$ 
have smaller $\Mstar$ compared to that for S05 relation, 
while the host galaxies in the case of $\sigma_{Z, {\rm int}} = 0.1$ have larger $\Mstar$. 
As a result, the mass PDF is less sensitive to the change of $\sigma_{Z, {\rm int}}$. 
This is because SFR and $\Zgal$ correlate stronger when $\Mstar$ is smaller in M11 relation. 
Once a galaxy sample is weighted with SFR, M11 relation makes $\Mstar$--$\Zgal$ relation steeper, 
and hence a difference in $\Zgal$ corresponds to smaller difference in $\Mstar$ with M11 relation 
compared to the case with S05 relation. 
However, we note that the predicted mass PDF is strongly dependent 
on the low-metallicity tail of M11 relation, which is still highly uncertain. 

\section{Discussions}
\label{sec:discussion}
We have predicted the metallicity and mass PDFs of GRB host galaxies, 
assuming empirical formulations of galaxy properties and the model of GRB rate 
in which GRBs occur only from low-metallicity stars ($< \Zcrit$: 12+log$_{10}$(O/H) = 8.2). 
Our results show that $>$ 50\% of GRB host galaxies can have $\Zgal > \Zcrit$, 
and high-metallicity galaxies (12+log$_{10}$(O/H) $\ga$ 8.8) 
may have significant contribution to cosmic GRB rate. 
This means that metallicities of GRB host galaxies 
may be systematically different from those of GRB progenitors, 
and the low-metallicity scenario can be reconciled 
with the observations of high-metallicity host galaxies of GRBs. 

For some GRBs, metallicities are measured 
at the positions of the bursts \citep[e.g.,][]{Modjaz:08a}, 
and the host galaxy of GRB 020819 has high-metallicity 
at the position of the burst \citep{Levesque:10d}. 
However, it should be noted that the positioning error 
of GRB 020819 is roughly 5 kpc \citep{Jakobsson:05b}, 
and there might be chemical inhomogeneity in the error circle. 
We need more precise localization of GRBs to draw robust conclusions, 
although it is also difficult to specify what precision is required. 
The required precision is dependent on the mixing process 
of inter-stellar medium which is not well understood. 

Although we have formulated $\epsilon_{\rm GRB}(\Zgal)$ motivated 
by the probable chemical inhomogeneity within each GRB host galaxy, 
similar formulation of $\epsilon_{\rm GRB}(\Zgal)$ 
may be obtained considering other effects 
(e.g. moderate low-metallicity preference 
of GRB occurrence without sharp metallicity cutoff). 
It is currently difficult to distinguish what effect constructs $\epsilon_{\rm GRB}(\Zgal)$. 

We have shown that multi-peak distribution of the metallicity of GRB host galaxies 
can be produced by a single population of GRB progenitors, 
when $\sigma_{Z, {\rm int}}$ positively correlates with $\Mstar$. 
If observed, the bimodality can be a clue to investigate the nature of GRB progenitors. 
If $\epsilon_{\rm GRB} > 0$ in high-metallicity galaxies 
is caused by the nature of GRB progenitors rather than properties of galaxies, 
there would be no effect of the $\sigma_{Z, {\rm int}}$--$\Mstar$ correlation. 

Although some results shown in this paper suffer from uncertainties 
about the properties of low-redshift galaxies, 
some important features of the results which we have discussed 
are not dependent on the detail of the modelings (see \S~\ref{sec:result}). 
However, we need to understand the actual metallicity distribution 
within young star forming galaxies to make reliable prediction of the exact metallicity PDF. 
More detailed study of the internal structure of galaxies 
requires different approach from that in this study, 
such as high-resolution hydrodynamic simulation 
and/or spatially resolved spectroscopic observation of large sample of galaxies, 
and we address this issue to future studies. 
Future development of our knowledge about galaxy properties 
would provide us with more robust predictions about GRB progenitors and their host galaxies. 

\section*{Acknowledgments}

We would like to thank an anonymous referee for his/her helpful comments. 
YN was supported by the Grant-in-Aid for JSPS Fellows. 

\bibliographystyle{mn2e}
\bibliography{reference_list}

\bsp

\label{lastpage}

\end{document}